\PassOptionsToPackage{hypertexnames=false}{hyperref}
\documentclass[pdflatex,sn-mathphys-num,iicol]{sn-jnl}
\usepackage[utf8]{inputenc}
\usepackage{graphicx,amsmath,amssymb,booktabs,array,tabularx}
\usepackage{placeins}
\usepackage{url}
\usepackage{needspace,balance}
\AtBeginDocument{\urlstyle{tt}}
\begin{document}
\title[Privacy-Aware Adaptive Client Selection for Federated RAG]{Privacy-Aware Adaptive Client Selection for Federated Retrieval-Augmented Generation}
\author[1]{\fnm{Ali} \sur{Shams Saryazd}}\email{alishams4d@gmail.com}
\author*[2]{\fnm{Abolfazl} \sur{Gandomi}}\email{abolfazl.gandomi@iau.ac.ir}
\affil[1]{\orgdiv{Department of Computer Engineering}, \orgname{Yazd University}, \orgaddress{\city{Yazd}, \country{Iran}}}
\affil[2]{\orgdiv{Department of Computer Engineering}, \orgname{Islamic Azad University, Yazd Branch}, \orgaddress{\city{Yazd}, \country{Iran}}}

\abstract{Federated retrieval-augmented generation retrieves evidence from distributed corpora, but broadcasting queries increases retrieval work and query exposure. We present Privacy-aware Adaptive Client Selection for Federated Retrieval-Augmented Generation (PACS-FedRAG), an online selector combining perturbed semantic profiles with evidence feedback, latency, payload cost, and exploration. In a ten-client simulation selecting three clients, the archived end-to-end benchmark reports evidence overlap of 0.372 and answer accuracy of 0.287. However, the retrieval-only evaluation reuses a full-broadcast reference with different fusion tie-breaking. Disjoint-client rank fusion also imposes an overlap ceiling of 0.375, limiting its relevance interpretation. Accuracy is below a fixed-label diagnostic of 0.645, and raw generator responses are unavailable for parser verification. Corrected baselines and new answer-level evaluation are therefore required to assess quality preservation. Selecting three clients reduces query recipients by 70\%; profile perturbation does not provide query or passage confidentiality.}


\maketitle

\section{Introduction}

Retrieval-augmented generation (RAG) combines parametric generation with non-parametric evidence retrieval, allowing a language model to condition its output on external documents \cite{ref1}. The architecture is particularly useful for knowledge-intensive question answering because the evidence store can be updated without retraining the generator. In many practical deployments, however, useful evidence is distributed across hospitals, laboratories, companies, or devices that cannot pool their complete document collections because of governance, ownership, confidentiality, or infrastructure constraints. Federated learning established a broader paradigm for collaboration without centralizing raw training data \cite{ref2,ref3}; federated RAG extends the same decentralization principle to retrieval and generation.

The Flower FedRAG example provides a reference pipeline for federated retrieval \cite{ref4}. It queries client-side FAISS indexes and combines returned candidates before generation. This architecture provides a useful basis for studying selective retrieval, but full broadcast requires retrieval work at every available source. Our implementation reuses Flower FedRAG components in a sequential simulation; it is not a deployment benchmark of independent organizations or a distributed Flower Message API experiment.

Full broadcast increases the number of query recipients and the aggregate local retrieval work. It also enlarges the candidate pool processed by the coordinator; the effect on latency depends on concurrency, network conditions, fusion, and generation. A query sent to an irrelevant source is still disclosed to that source. We therefore distinguish the number of recipients, a data-minimization measure, from formal query confidentiality. The system problem is to allocate a limited retrieval budget while retaining useful evidence, not simply to reduce the number of contacted clients.

Source selection is a classical problem in federated information retrieval \cite{ref5}. RAGRoute applies a lightweight neural classifier to source selection for retrieval-augmented generation \cite{ref6}. Client selection itself is therefore not the novelty claimed here. PACS-FedRAG investigates a complementary design: query-dependent semantic matching combined with online evidence and system feedback, without task-specific router training.

RAGRoute is retained as closely related work, but is not implemented or evaluated in this study; no empirical superiority over it is claimed.

PACS-FedRAG (Privacy-aware Adaptive Client Selection for Federated RAG) scores clients using a perturbed corpus centroid, historical evidence contribution, measured retrieval latency, estimated payload cost, and an exploration bonus. It selects a fixed number of available clients, combines their candidates using reciprocal rank fusion (RRF), and updates client statistics after retrieval. This design provides a lightweight routing policy for a coordinator that cannot inspect complete remote corpora in its intended deployment. Its current evaluation isolates the selection policy in a local simulation.

The contributions are as follows:

\begin{itemize}

\item An online, fixed-budget selector that combines semantic relevance with historical evidence contribution, retrieval latency, payload cost, and exploration, without fitting a separate supervised router.

\item A Gaussian-perturbed profile mechanism with an explicit replace-one adjacency model and sensitivity calculation, separating profile protection from passage disclosure and query-recipient reduction.

\item A comparison against full broadcast, random selection, round-robin selection, and static perturbed-profile selection on two biomedical question sets, with fixed retrieval budgets and published per-query records.

\item A retrieval-only privacy-budget sweep and component analysis with per-query records and calibration provenance, with explicit limitations on baseline compatibility and overlap-metric validity, and an assessment of deterministic rerun consistency.

\end{itemize}

The paper does not claim end-to-end confidential inference, a new differential privacy mechanism, or an optimal routing policy. Selected passages are intentionally disclosed to the coordinator, and queries are visible to selected clients. The intended privacy scope concerns released profiles under the assumptions in Section 3.3. The publicly seeded simulation is not a production privacy guarantee.

\section{Related Work}

\subsection{Retrieval-Augmented Generation}

Lewis et al. introduced RAG as a sequence-to-sequence generation framework that combines a pretrained generator with retrieved passages \cite{ref1}. Subsequent RAG systems commonly use dense embeddings, approximate nearest-neighbor search, candidate reranking, and evidence-conditioned generation. FAISS provides efficient vector indexing and similarity search primitives used by many such pipelines \cite{ref7}. For medical RAG evaluation, MIRAGE/MedRAG unifies several QA datasets and four corpora and reports both retrieval and answer-level outcomes \cite{ref8}. These components motivate our use of the existing Flower pipeline instead of constructing an unrelated system.

\subsection{Federated RAG and Confidential Retrieval}

Federated RAG retains source documents at participating data providers while coordinating retrieval and generation. C-FedRAG places orchestration and inference inside a confidential-computing environment to protect intermediate results from an over-curious orchestrator \cite{ref9}. FRAG studies secure and efficient federated vector-database management and cryptographic protocols for collaborative retrieval \cite{ref10}. FedE4RAG focuses on federated embedding learning, knowledge distillation, and homomorphic protection of exchanged model information \cite{ref11}. Other work combines federated fine-tuning and RAG for domain-specific language models \cite{ref12}. A systematic mapping study identifies privacy-preserving retrieval, cross-client heterogeneity, and limited evaluation practice as central open concerns \cite{ref13}.

These systems span different privacy goals. Keeping raw corpora local prevents wholesale data centralization but does not by itself guarantee that retrieved passages, semantic profiles, model updates, or queries reveal no sensitive information. C-FedRAG explicitly calls for threat modeling and considers differential privacy and trusted execution environments as complementary technologies \cite{ref9}. PACS-FedRAG therefore states a limited threat model and applies formal perturbation only to the profile used for routing.

\subsection{Client and Source Selection}

Federated learning uses client selection to manage statistical utility, system heterogeneity, and training speed. FedAvg samples clients and averages local model updates \cite{ref3}. Oort preferentially selects clients that offer high statistical utility while accounting for execution speed \cite{ref14}. Power-of-choice selection shows that non-uniform client choice can accelerate federated optimization under heterogeneous data \cite{ref15}. Our problem differs because clients do not train a shared model in each query round; they retrieve evidence. Nevertheless, the underlying idea---balancing utility, cost, and exploration---transfers naturally.

In federated search, resource selection estimates which collections are useful for a query \cite{ref5}. RAGRoute is the closest related source-routing approach: it uses a trained neural classifier rather than the heuristic online score studied here \cite{ref6}. PACS-FedRAG combines profile similarity with observed evidence contribution and system costs, and perturbs routing profiles. These are design distinctions, not measured advantages over RAGRoute. The archived experiments include only non-neural selectors, and the validity limitations in Sections 4.6--4.10 constrain their interpretation. A controlled comparison with RAGRoute would require common corpus partitions, retrieval settings, evaluation queries, and a stated router-training protocol.

Routing also introduces integrity risks. A client that advertises a misleading semantic profile can attract queries and return irrelevant or poisoned evidence \cite{ref16}. We assume honest clients and an honest-but-curious coordinator. Evidence feedback may reduce a client's score after unhelpful returns, but this is not a Byzantine-robust defense and has not been evaluated as a poisoning detector.

\subsection{Differential Privacy}

Differential privacy (DP) limits changes in the output distribution between neighboring datasets \cite{ref17,ref18}. Here, neighboring corpora have the same public size and differ in one indexed record. The retrieval-only implementation uses the analytically calibrated Gaussian mechanism \cite{ref19} for a bounded-sensitivity centroid. This calibration provenance does not extend retrospectively to the earlier end-to-end run. Releasing a new independently randomized profile consumes privacy budget; reusing the same released profile for computations based only on that profile is post-processing. Composition would be required for repeated releases \cite{ref20,ref21}.

\section{Methodology}

\subsection{System Model}

Let C = \{1, \ldots{}, N\} denote a federation of N clients. Client i stores a corpus D\textsubscript{i} with n\textsubscript{i} indexed records and a local FAISS index. A record is an indexed text chunk; protection for a complete source document containing multiple chunks would require separate group-privacy accounting. Queries arrive sequentially. A coordinator selects S\textsubscript{t} from the available clients A\textsubscript{t}, with \textbar{}S\textsubscript{t}\textbar{} = min(m, \textbar{}A\textsubscript{t}\textbar{}), and provides returned evidence to a generator. All clients use a common embedding model E. This is the intended logical architecture; the evaluated implementation keeps the client indexes in one simulation process.

Each selected client returns its local top-k documents and retrieval scores. The server merges the candidates, retains a global top-k list, and augments the generation prompt. The selection objective is multi-criteria: preserve retrieval and answer quality while reducing contacted clients, latency, communication volume, and query exposure.

\begin{equation}
\begin{aligned}S_t^{\star}&=\operatorname*{arg\,max}_{S\subseteq A_t,\ |S|=\min(m,|A_t|)}J_t(S),\\J_t(S)&=\mathbb{E}[Q(q_t,S)]-\lambda_L L(q_t,S)\\&\quad-\lambda_C C(q_t,S)-\lambda_X X(S).\end{aligned}\tag{1}\label{eq:1}
\end{equation}

In Eq. (1), Q denotes downstream answer utility, L denotes latency, C denotes communication cost, and X denotes the number of query recipients. The equation motivates the design rather than a solved optimization problem. Under a fixed budget and full availability, X is constant across eligible subsets, so the score distinguishes clients by relevance and system feedback. The per-client latency penalty is a surrogate: parallel response time may instead be determined by the slowest selected client. PACS-FedRAG makes no optimality or regret guarantee for this objective. Figure 1 illustrates the logical routing workflow.

\begin{figure*}[t]\centering
\includegraphics[width=\textwidth]{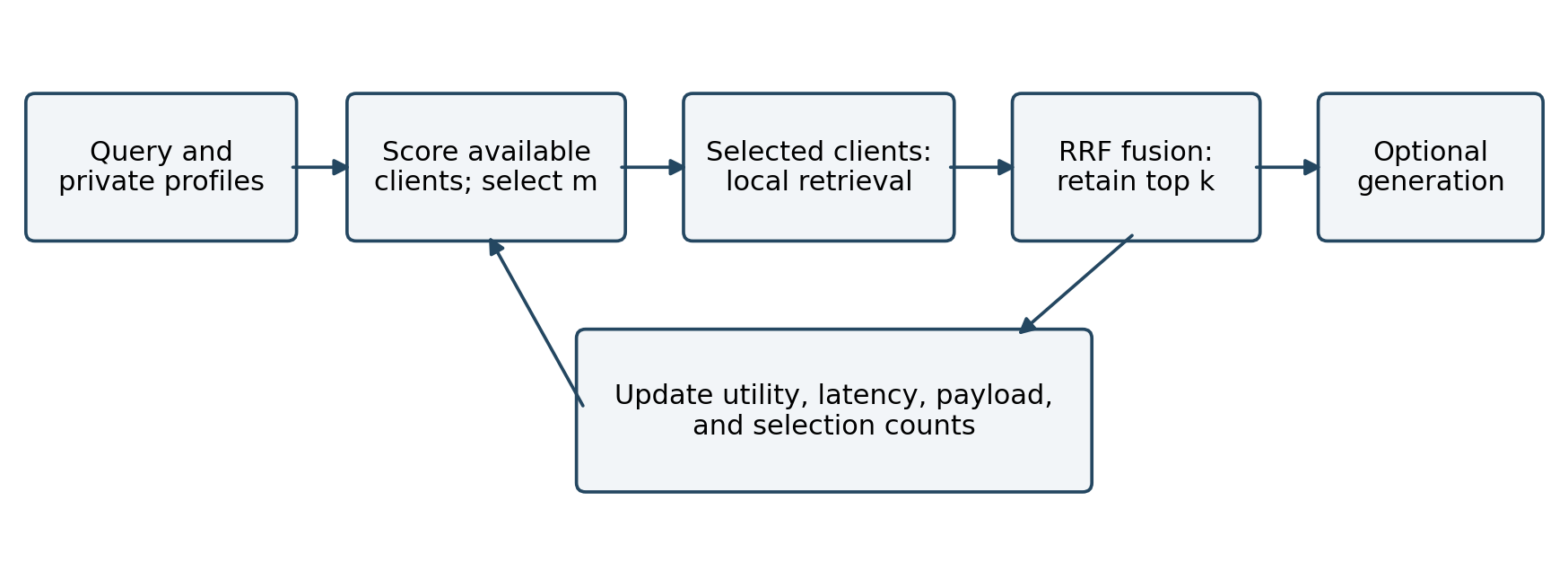}
\caption{Logical PACS-FedRAG workflow. Evidence feedback follows fusion; generation is optional and disabled in the retrieval-only experiments. There is no separate neural reranker. Privacy qualifications appear in Section 3.3.}\label{fig:1}
\end{figure*}

\subsection{Threat Model and Privacy Scope}

The intended coordinator is honest-but-curious and clients are honest. Authenticated, encrypted transport is an operational assumption, not a property tested by the local simulation. Complete corpora and indexes remain client-side in the intended architecture, but are loaded locally by the experimental driver. Selected passages are deliberately returned for generation and lie outside the profile privacy mechanism. The coordinator and selected clients observe query content. Recipient reduction measures less dissemination, not concealment from those parties.

The protected contribution is one indexed record in a released client profile. Adjacency means replacing one bounded record in a corpus of fixed public size. A fixed profile can be reused for profile-only computations without an additional release, whereas regenerating noisy profiles requires composition. A server transcript that also contains retrieved passages and evidence-derived feedback is not covered by a profile-only DP claim. The simulation generates profiles separately for each seed; those runs do not constitute a single accounted private deployment.

\subsection{Private Semantic Client Profiles}

Client i L2-normalizes each nonzero record embedding, clips it to norm at most C\textsubscript{e}, and computes its mean \ensuremath{\mu}\textsubscript{i}. Projecting the mean onto an L2 ball of radius C\textsubscript{p} gives the following definition, including the zero-vector case:

\begin{equation}
\bar\mu_i=\begin{cases}\mu_i\min\!\left(1,\dfrac{C_p}{\|\mu_i\|_2}\right),&\|\mu_i\|_2>0,\\0,&\|\mu_i\|_2=0.\end{cases}\tag{2}\label{eq:2}
\end{equation}

The profile is perturbed before normalization for cosine scoring:

\begin{equation}
\widetilde p_i=\bar\mu_i+\mathcal{N}(0,\sigma_i^2 I_d)\tag{3}\label{eq:3}
\end{equation}

In Eq. (3), \ensuremath{\sigma}\textsubscript{i} is the absolute per-coordinate Gaussian noise standard deviation. Replacing one embedding clipped to norm C\textsubscript{e} changes the mean by at most \ensuremath{\Delta}\textsubscript{i}=2C\textsubscript{e}/n\textsubscript{i} under fixed-size replace-one adjacency. Euclidean projection onto the profile ball is non-expansive, so it does not increase sensitivity. With C\textsubscript{e}=C\textsubscript{p}=1, the mean is already inside the profile ball; projection is retained as a safeguard. The retrieval-only implementation solves the analytic Gaussian calibration equation \cite{ref19} for each public client size n\textsubscript{i}. Here \ensuremath{\Phi} is the standard normal cumulative distribution function:

\begin{equation}
\begin{aligned}\delta(\sigma_i)&=\Phi\!\left(\dfrac{\Delta_i}{2\sigma_i}-\dfrac{\varepsilon\sigma_i}{\Delta_i}\right)\\&\quad-e^{\varepsilon}\Phi\!\left(-\dfrac{\Delta_i}{2\sigma_i}-\dfrac{\varepsilon\sigma_i}{\Delta_i}\right),\\\Delta_i&=\dfrac{2C_e}{n_i}.\end{aligned}\tag{4}\label{eq:4}
\end{equation}

Client corpus sizes are public simulation metadata, and normalizing the noisy vector is post-processing. The retrieval-only configuration uses C\textsubscript{e}=C\textsubscript{p}=1 and \ensuremath{\delta}=10\textsuperscript{-}\textsuperscript{5}. At \ensuremath{\varepsilon}=4, recorded \ensuremath{\sigma}\textsubscript{i} values have mean 0.005411, minimum 0.005029, and maximum 0.005705 across ten clients; numerical substitution into Eq. (4) gives a maximum absolute \ensuremath{\delta} error of approximately 7.9\ensuremath{\times}10\textsuperscript{-}\textsuperscript{1}\textsuperscript{3}. These values describe the retrieval-only profiles, not the archived Table 3 profiles. One centroid is used per client. Seeds 13, 37, and 73 provide reproducible simulation, not production-private randomness. An operational implementation would require fresh secure randomness and appropriate privacy accounting; its random-number implementation and finite-precision behavior have not been independently audited here. The profile mechanism does not cover queries, returned passages, or the complete evidence-feedback transcript.

\subsection{Adaptive Client Score}

For query q\textsubscript{t}, the server computes a normalized query embedding z\textsubscript{t} and private semantic similarity s\textsubscript{i},\textsubscript{t}. The denominator guard and zero-vector convention avoid undefined normalization:

\begin{equation}
\begin{gathered}z_t=\dfrac{E(q_t)}{\max(\|E(q_t)\|_2,10^{-10})},\\s_{i,t}=\begin{cases}\dfrac{z_t^{\mathsf T}\widetilde p_i}{\|z_t\|_2\|\widetilde p_i\|_2},&\|z_t\|_2\|\widetilde p_i\|_2>0,\\0,&\text{otherwise.}\end{cases}\end{gathered}\tag{5}\label{eq:5}
\end{equation}

For each client the coordinator maintains utility u\textsubscript{i},\textsubscript{t}, smoothed retrieval latency \ensuremath{\ell}\textsubscript{i},\textsubscript{t}, smoothed response payload c\textsubscript{i},\textsubscript{t}, and selection count n\textsubscript{i},\textsubscript{t}. Utility, latency, and payload are min-max normalized across available clients; a zero range maps to 0.5 in the implementation. Statistics start at zero. The following upper-confidence-style bonus favors less frequently selected clients, but is used heuristically rather than as a confidence bound with a regret proof:

\begin{equation}
b_{i,t}=\sqrt{\dfrac{2\ln(t+1)}{n_{i,t}+1}}\tag{6}\label{eq:6}
\end{equation}

The final score is:

\begin{equation}
\begin{aligned}
g_{i,t}&=\alpha s_{i,t}+\beta\widehat u_{i,t}\\[2pt]
&\quad-\gamma\widehat{\ell}_{i,t}^{\;A}-\eta\widehat c_{i,t}+\xi b_{i,t}
\end{aligned}\tag{7}\label{eq:7}
\end{equation}

The coordinator selects min(m, \textbar{}A\textsubscript{t}\textbar{}) clients with the largest scores. The weights are \ensuremath{\alpha}=0.45, \ensuremath{\beta}=0.30, \ensuremath{\gamma}=0.15, \ensuremath{\eta}=0.05, and \ensuremath{\xi}=0.05. They are heuristic settings. Superscript A denotes normalization of stored smoothed latency over available clients. The implementation uses a zero-based query index and log(t+2), equivalent to the one-based convention in Eq. (6). No optimality, universal superiority, or hyperparameter robustness is inferred from this configuration.

\subsection{Retrieval, Fusion, and Reward}

Each selected client returns k local candidates. RRF \cite{ref22} combines ranks without requiring a common distance scale, although the evaluated clients use a common embedding model. Let rank\textsubscript{i},\textsubscript{t}(d) be the one-based local rank of document d in client i's returned set R\textsubscript{i},\textsubscript{t}. Its fusion score is:

\begin{equation}
\operatorname{RRF}(d)=\sum_{i\in S_t:\ d\in R_{i,t}}\dfrac{1}{\kappa+\operatorname{rank}_{i,t}(d)}\tag{8}\label{eq:8}
\end{equation}

The constant \ensuremath{\kappa} is 60 and the highest-ranked k fused candidates form G\textsubscript{t}. No additional neural reranker is applied. The archived end-to-end implementation resolves equal scores by candidate insertion order; the retrieval-only implementation resolves them by ascending string document ID. These distinct protocols were both labeled ``legacy\_rrf'' in the artifacts. Section 4.6 explains why the distinction matters for the reused Full baseline. For a nonempty evidence set, the client reward is:

\begin{equation}
\begin{aligned}r_{i,t}&=\omega\dfrac{|R_{i,t}\cap G_t|}{|G_t|}\\&\quad+(1-\omega)\bar v_{i,t}-\tau\widehat{\ell}_{i,t}^{\;S}\end{aligned}\tag{9}\label{eq:9}
\end{equation}

The first evidence term is the fraction of G\textsubscript{t} supplied by client i. The second is the mean reciprocal-rank contribution of that client's surviving records, without separate min--max normalization:

\begin{gather}
H_{i,t}=R_{i,t}\cap G_t,\tag{9a}\label{eq:9a}\\*[3pt]
\bar v_{i,t}=\begin{cases}
\displaystyle\frac{\sum_{d\in H_{i,t}}\! (\kappa+\operatorname{rank}_{i,t}(d))^{-1}}{|H_{i,t}|},&|H_{i,t}|>0,\\[3pt]
0,&|H_{i,t}|=0.
\end{cases}\notag
\end{gather}

If G\textsubscript{t} is empty, both evidence terms are defined as zero. We use \ensuremath{\omega}=0.5 and \ensuremath{\tau}=0.1. In Eq. (9), superscript S denotes normalization of current retrieval latency over selected clients, whereas Eq. (7) normalizes stored smoothed latency over available clients. Latency therefore enters both historical utility and the explicit score penalty. The utility update is:

\begin{equation}
\begin{gathered}u_{i,t+1}=(1-\rho)u_{i,t}+\rho r_{i,t},\\\rho\in(0,1]\end{gathered}\tag{10}\label{eq:10}
\end{equation}

The exponentially weighted moving average (EWMA) uses \ensuremath{\rho}=0.2; latency and payload statistics are smoothed with the same rate. Unselected clients retain their previous state. Reward reflects survival under fusion, not verified relevance or causal improvement in answer quality. It is observed only for selected clients and can reinforce earlier selections. Exploration mitigates exclusion heuristically but does not establish unbiased feedback. Algorithm 1 summarizes the implemented sequence.

\par\Needspace{6\baselineskip}\medskip\noindent\textbf{Algorithm 1. PACS-FedRAG selection and feedback}\par\nobreak\smallskip

Input: query q\textsubscript{t}; available clients A\textsubscript{t}; profiles; budget m; scoring weights; EWMA rate \ensuremath{\rho}. Output: selected clients, evidence, updated state, and an answer when generation is enabled.

\begin{enumerate}

\item Normalize the query embedding, with a zero-norm guard, and compute profile similarities.

\item Normalize stored utility, latency, and response-payload values over available clients.

\item Compute the exploration bonus and composite score using Eqs. (6) and (7).

\item Sort scores and select min(m, \textbar{}A\textsubscript{t}\textbar{}) clients.

\item Retrieve eight local candidates from each selected client index.

\item Fuse candidates with the explicitly chosen RRF tie-breaking protocol (\ensuremath{\kappa}=60); retain eight records.

\item Compute the evidence reward and update utility, latency, response payload, and selection counts.

\item If enabled, generate an answer from the fused context; log the prediction and stage timings.

\end{enumerate}

\subsection{Complexity and Communication}

Computing profile similarities and client statistics costs O(Nd) per query and stores O(Nd) profile coordinates plus O(N) state. The current implementation sorts all scores, costing O(N log N); a bounded heap could reduce selection to O(N log m), but is not used here. Only m indexes are searched instead of N, and up to mk candidates replace up to Nk. The recipient ratio is m/N and the ideal candidate-count reduction is 1\ensuremath{-}m/N. These arithmetic reductions follow from the budget and are shared by all fixed-m selectors. Actual payload and latency depend on returned text length, scheduling, transport, and generation. The experiment measures the sequential simulation rather than parallel cluster speedup.

\section{Experiments}

\subsection{Research Questions}

The evaluation separates the earlier end-to-end benchmark from the retrieval-only validation and addresses four questions:

\begin{itemize}

\item RQ1: What answer accuracy and overlap with full-broadcast evidence are observed when three of ten clients are selected?

\item RQ2: How does PACS-FedRAG compare with random, round-robin, and static perturbed-profile selection at the same budget?

\item RQ3: How do analytically calibrated profile-noise settings affect selected-client agreement, evidence overlap, participation uniformity, and communication volume?

\item RQ4: How are evidence overlap, participation uniformity, and communication volume associated with removing historical utility, latency components, or exploration from the selector?

\end{itemize}

\subsection{Datasets and Federation Construction}

The experiment uses the StatPearls and Textbooks retrieval corpora and the PubMedQA and BioASQ question sets through the Flower FedRAG/MIRAGE workflow \cite{ref4,ref8}. PubMedQA originates from biomedical research questions \cite{ref23}; BioASQ provides biomedical question-answering tasks \cite{ref24}. The evaluated data are the option- based versions loaded by the repository, not every task or original answer format in those benchmarks. Table 1 lists the subsets used.

The federation contains N=10 simulated clients: five per retrieval corpus. Each corpus contributes 2,000 indexed chunks. The repository assigns chunk identifiers to clients by a seeded SHA-256 hash modulo five. This creates coarse source heterogeneity, but is not a controlled topic-skew or realistic institutional partition. Chunks from one original source document need not remain in one silo. The driver evaluates the first 100 loaded questions per dataset, using the same order across methods and seeds. Thus there are 200 distinct question identifiers, rather than 600 independent questions across the three runs.

The partition seed is 42: a chunk d is assigned within its corpus to integer(SHA-256("42\_" + d)) modulo five. This chunk-level rule does not keep all chunks from one source document together. The scorer is initialized anew for each dataset and seed.

\begin{table*}[t]\centering
\caption{Datasets and corpus subsets in the ten-client simulation.}\label{tab:1}
\small\setlength{\tabcolsep}{3pt}\renewcommand{\arraystretch}{1.12}
\newdimen\usablewidth\usablewidth=\dimexpr\textwidth-24pt\relax
\begin{tabular}{@{}>{\raggedright\arraybackslash}p{0.2091\usablewidth}>{\raggedright\arraybackslash}p{0.2182\usablewidth}>{\raggedright\arraybackslash}p{0.2091\usablewidth}>{\raggedright\arraybackslash}p{0.1091\usablewidth}>{\raggedright\arraybackslash}p{0.2545\usablewidth}@{}}\toprule

\textbf{Dataset / corpus} & \textbf{Role} & \textbf{Items used} & \textbf{Clients} & \textbf{Partition}\\\midrule

StatPearls & Retrieval corpus & 2,000 chunks & 5 & Hash by chunk ID\\

Textbooks & Retrieval corpus & 2,000 chunks & 5 & Hash by chunk ID\\

PubMedQA & QA benchmark & 100 questions & Server & Evaluation only\\

BioASQ & QA benchmark & 100 questions & Server & Evaluation only\\

\bottomrule\end{tabular}
\end{table*}

\subsection{Implementation Details}

The implementation reuses the Flower FedRAG retriever and MIRAGE question loader in a standalone sequential driver. It builds FAISS IndexIVFFlat indexes with L2 distance; each selected index returns eight candidates, and RRF (\ensuremath{\kappa}=60) retains eight records. The embedding model is sentence-transformers/all-MiniLM-L6-v2 at the pinned revision recorded in the public run manifests, producing 384-dimensional vectors. Raw vectors are supplied to the L2 FAISS index, whereas normalized vectors are used for cosine profile scoring. The earlier end-to-end benchmark uses Qwen2.5-7B-Instruct Q4\_K\_M. The privacy and component evaluations are retrieval-only: they make no generator call and therefore report neither answer accuracy nor generation-inclusive end-to-end latency.

Both protocols use C\textsubscript{e}=C\textsubscript{p}=1, \ensuremath{\delta}=10\textsuperscript{-}\textsuperscript{5}, and m=3 of N=10 clients. The retrieval-only sweep evaluates \ensuremath{\varepsilon}\ensuremath{\in}\{1,4,\ensuremath{\infty}\}, where \ensuremath{\infty} denotes an unperturbed centroid, with simulation seeds 13, 37, and 73 over 100 PubMedQA and 100 BioASQ questions. The scoring weights are \ensuremath{\alpha}=0.45, \ensuremath{\beta}=0.30, \ensuremath{\gamma}=0.15, \ensuremath{\eta}=0.05, and \ensuremath{\xi}=0.05, with EWMA rate \ensuremath{\rho}=0.2. Retrieval-latency penalties use a fixed calibration profile paired with a verified workload manifest. Exact artifact versions and SHA-256 digests are recorded in the public run manifests.

The index construction uses nlist=max(1, min(100, floor(\ensuremath{\surd}n\textsubscript{i}))) and nprobe=min(10, nlist). The recorded retrieval-only environment is macOS 26.6.2 on arm64, Python 3.12.3, NumPy 1.26.4, FAISS 1.14.3 with eight threads, PyTorch 2.13.0, and sentence-transformers 5.6.1. These are manifest-reported versions; an exact processor model is not recorded. Calibration uses the fixed 200-question workload with 20 warmup and 100 measurement queries per client. Calibrated latency used by the scorer is distinct from observed wall-clock retrieval latency.

The earlier end-to-end run uses a historical \ensuremath{\varepsilon}=4 profile setting without the per-client analytic provenance available for v6. The archived classical Gaussian formula is not automatically invalid at \ensuremath{\varepsilon}=4: with \ensuremath{\delta}=10\textsuperscript{-}\textsuperscript{5} it gives \ensuremath{\sigma}/\ensuremath{\Delta}\ensuremath{\approx}1.2112, whose analytic \ensuremath{\delta} is approximately 6.8\ensuremath{\times}10\textsuperscript{-}\textsuperscript{7}. This mathematical check does not recover missing run-specific profile provenance or turn publicly seeded simulation into a production privacy guarantee.

\begin{table*}[t]\centering
\caption{Implementation configuration and protocol-specific provenance.}\label{tab:2}
\small\setlength{\tabcolsep}{3pt}\renewcommand{\arraystretch}{1.12}
\newdimen\usablewidth\usablewidth=\dimexpr\textwidth-6pt\relax
\begin{tabular}{@{}>{\raggedright\arraybackslash}p{0.2636\usablewidth}>{\raggedright\arraybackslash}p{0.7364\usablewidth}@{}}\toprule

\textbf{Component} & \textbf{Setting}\\\midrule

Framework / budget & Flower FedRAG components; sequential simulation; N=10, m=3\\

Corpora / questions & StatPearls and Textbooks; 100 PubMedQA + 100 BioASQ questions\\

Local / global top-k & 8 / 8\\

Fusion & RRF, \ensuremath{\kappa}=60; Table 3: insertion-order ties; v6 selective: ascending document-ID ties\\

Embedding & all-MiniLM-L6-v2; 384 dimensions\\

Generator & Qwen2.5-7B-Instruct Q4\_K\_M for Table 3; temperature 0, max tokens 10; disabled for v6\\

Profile privacy, v6 & Analytic Gaussian; C\textsubscript{e}=C\textsubscript{p}=1; \ensuremath{\delta}=10\textsuperscript{-}\textsuperscript{5}; \ensuremath{\varepsilon}\ensuremath{\in}\{1,4,\ensuremath{\infty}\}; deterministic simulation\\

Profile privacy, Table 3 & Historical \ensuremath{\varepsilon}=4 setting; per-client analytic provenance not established\\

Simulation seeds & 13, 37, 73\\

Full reference in v6 & 600 archived Table 3 references reused; 0 recomputed; tie-breaking incompatibility unresolved\\

\bottomrule\end{tabular}
\end{table*}

\subsection{Baselines}

Full queries all ten client indexes and serves as the broadcast reference. Random-3 samples three available clients uniformly; Round-Robin-3 cycles through clients without query information. Static-DP-3 selects the three highest profile similarities without evidence feedback. PACS-DP-3 uses the complete score. All selective methods have the same client budget. RAGRoute \cite{ref6} is discussed as closely related work but was not implemented or evaluated. Consequently, these experiments do not compare PACS-FedRAG with supervised neural routing or support a claim of superiority over RAGRoute. No-RAG is not an evaluated condition.

\subsection{Metrics}

Answer accuracy is the fraction of parsed option labels matching the reference label, including missing predictions as incorrect, and is available only for the archived end-to-end benchmark. Evidence overlap@8 is the fraction of the reference Full output's eight record identifiers also returned by a selective method for the same dataset, question, and seed. It is not gold-passage recall. Tables 4 and 5 use archived Full outputs with different fusion tie-breaking, so their metric is explicitly labeled archived-reference overlap. Selected-client routing overlap is Jaccard similarity with the \ensuremath{\varepsilon}=\ensuremath{\infty} route. Jain's index \cite{ref25} uses all ten client selection counts, including zeros. Reported bytes count UTF-8 query text sent to selected clients plus returned text; they omit record IDs, scores, serialization, headers, and transport overhead. They are estimated text-payload volume, not measured network traffic. The reward's response-payload statistic excludes sent query text.

\begin{equation}
\operatorname{ClientReduction}(t)=1-\dfrac{|S_t|}{N}\tag{11}\label{eq:11}
\end{equation}

\begin{equation}
\operatorname{Jain}=\dfrac{\left(\sum_{i=1}^{N}f_i\right)^2}{N\sum_{i=1}^{N}f_i^2}\tag{12}\label{eq:12}
\end{equation}

Here f\textsubscript{i} is client i's selection count and at least one selection is assumed. Jain's index describes participation concentration, not equal relevance or equitable answer quality. Maximum uniformity need not be desirable under heterogeneous corpora. The complete ten-client count vector is used. Client reduction in Eq. (11) uses the total federation size N; it is 70\% for the evaluated fully available ten-client setting.

\subsection{Main Results}

\begin{table*}[t]\centering
\caption{Archived end-to-end benchmark, pooled over 200 distinct questions and three seeds. \ensuremath{\pm} is population SD across seed means (ddof=0). Latency includes generation. Overlap uses archived Full; bytes are total query-plus-returned-text payload. \ensuremath{\varepsilon}=4 labels denote the historical setting, not v6 analytic provenance.}\label{tab:3}
\small\setlength{\tabcolsep}{3pt}\renewcommand{\arraystretch}{1.12}
\newdimen\usablewidth\usablewidth=\dimexpr\textwidth-30pt\relax
\begin{tabular}{@{}>{\raggedright\arraybackslash}p{0.2386\usablewidth}>{\raggedright\arraybackslash}p{0.1686\usablewidth}>{\raggedright\arraybackslash}p{0.1705\usablewidth}>{\raggedright\arraybackslash}p{0.1004\usablewidth}>{\raggedright\arraybackslash}p{0.2027\usablewidth}>{\raggedright\arraybackslash}p{0.1193\usablewidth}@{}}\toprule

\textbf{Method} & \textbf{Accuracy} & \textbf{Overlap@8} & \textbf{Clients} & \textbf{E2E ms} & \textbf{Text bytes}\\\midrule

Full & 0.287\ensuremath{\pm}0.002 & 1.000\ensuremath{\pm}0.000 & 10 & 7114.4\ensuremath{\pm}196.9 & 48,299\\

Random-3 & 0.293\ensuremath{\pm}0.006 & 0.298\ensuremath{\pm}0.007 & 3 & 8017.9\ensuremath{\pm}351.1 & 14,310\\

Round-Robin-3 & 0.295\ensuremath{\pm}0.011 & 0.300\ensuremath{\pm}0.000 & 3 & 8126.1\ensuremath{\pm}628.8 & 14,448\\

Static-DP-3 (\ensuremath{\varepsilon}=4) & 0.280\ensuremath{\pm}0.004 & 0.362\ensuremath{\pm}0.001 & 3 & 5900.1\ensuremath{\pm}36.8 & 10,652\\

PACS-DP-3 (\ensuremath{\varepsilon}=4) & 0.287\ensuremath{\pm}0.002 & 0.372\ensuremath{\pm}0.001 & 3 & 5492.0\ensuremath{\pm}385.0 & 9,878\\

\bottomrule\end{tabular}
\end{table*}

The archived records reproduce the displayed accuracy, overlap, latency, and payload summaries. PACS-DP-3 and Full have equal pooled accuracy to the shown precision, but this is not evidence of noninferiority. Full's final eight records always come from statpearls\_0 through statpearls\_4 and textbooks\_0 through textbooks\_2. With disjoint chunk IDs, all local rank-1 records tie under RRF; stable insertion order favors these eight clients. For every archived selective Table 3 record, and for the v6 privacy-sweep and component records examined, overlap with archived Full satisfies:

\begin{equation}
\begin{gathered}T=\{\texttt{textbooks\_3},\texttt{textbooks\_4}\},\\\operatorname{Overlap@8}(S)=\dfrac{3-|S\cap T|}{8}\\\leq\dfrac{3}{8}=0.375.\end{gathered}\tag{13}\label{eq:13}
\end{equation}

This identity makes 0.375 a structural ceiling and 0.300 the expected overlap of uniform Random-3 in this setting. Near-ceiling PACS overlap mostly indicates avoidance of textbooks\_3 and textbooks\_4, not verified relevance. Separately, the retrieval-only runner resolves ties lexicographically, yet its protocol check compares only the common fusion label and other settings. The v6 manifest records all 600 Full references as reused and none recomputed. Consequently, Tables 4 and 5 compare selective outputs with a different-protocol Full reference; they are retained as archived-reference diagnostics pending recomputation.

Answer-level diagnostics also limit interpretation. Among the 200 unique questions, PubMedQA has 58 A, 32 B, and 10 C labels; BioASQ has 71 A and 29 B labels. Always predicting A would score 129/200=0.645 on this subset. This is a descriptive fixed-label sanity check computed from the evaluation labels, not a trained baseline or an estimate for unseen data. Archived PACS accuracy is 0.247 on PubMedQA and 0.327 on BioASQ; Full is 0.210 and 0.363, respectively. Pooled equality masks these differences. The stored correctness flags agree with the stored parsed labels, but they cannot validate the original generated text.

The inspected end-to-end parser accepts a standalone A--D token and also a first-character fallback; ordinary prose can therefore become an option label. The PubMedQA prompt mixes yes/no/maybe and letter instructions, and generation is capped at ten tokens. Raw responses and finish reasons were not retained, so the contribution of parsing, truncation, prompting, retrieval, or model behavior to the low accuracy cannot be isolated retrospectively. Table 3 is preserved as an archived observation; validated parsing, retained raw outputs, and a same-settings no-RAG control require a new end-to-end run.

\subsection{Privacy Budget Analysis}

\begin{table*}[t]\centering
\caption{Archived retrieval-only privacy sweep: 200 questions \ensuremath{\times} three seeds. Means\ensuremath{\pm}sample SD across seed means (ddof=1); \ensuremath{\infty} means unperturbed profiles. Generation is disabled. The Full reference uses incompatible historical tie-breaking; overlap is not a same-protocol quality comparison.}\label{tab:4}
\small\setlength{\tabcolsep}{3pt}\renewcommand{\arraystretch}{1.12}
\newdimen\usablewidth\usablewidth=\dimexpr\textwidth-24pt\relax
\begin{tabular}{@{}>{\raggedright\arraybackslash}p{0.0727\usablewidth}>{\raggedright\arraybackslash}p{0.2727\usablewidth}>{\raggedright\arraybackslash}p{0.2727\usablewidth}>{\raggedright\arraybackslash}p{0.1909\usablewidth}>{\raggedright\arraybackslash}p{0.1909\usablewidth}@{}}\toprule

\textbf{\ensuremath{\varepsilon}} & \textbf{Overlap@8 vs archived Full} & \textbf{Routing Jaccard vs \ensuremath{\varepsilon}=\ensuremath{\infty}} & \textbf{Jain index} & \textbf{Total text bytes}\\\midrule

\ensuremath{\infty} & 0.351\ensuremath{\pm}0.000 & 1.000\ensuremath{\pm}0.000 & 0.493\ensuremath{\pm}0.000 & 12,330\\

4 & 0.358\ensuremath{\pm}0.003 & 0.706\ensuremath{\pm}0.010 & 0.479\ensuremath{\pm}0.005 & 12,425\\

1 & 0.351\ensuremath{\pm}0.003 & 0.571\ensuremath{\pm}0.017 & 0.475\ensuremath{\pm}0.017 & 13,283\\

\bottomrule\end{tabular}
\end{table*}

At \ensuremath{\varepsilon}=4 and \ensuremath{\varepsilon}=1, selected-client Jaccard overlap with the unperturbed route is 0.706\ensuremath{\pm}0.010 and 0.571\ensuremath{\pm}0.017, respectively. Thus the profile-noise setting changes routing on this fixed workload. Archived-reference evidence overlap is nonmonotonic, but it is constrained by Eq. (13) and by the fusion mismatch; it cannot establish that noise improves or preserves retrieval utility. The reported total text-payload means are 12,330, 12,425, and 13,283 bytes for \ensuremath{\varepsilon}=\ensuremath{\infty}, 4, and 1. Returned-text means alone are approximately 12,106, 12,200, and 13,059 bytes. Matching deterministic fields across the two stored runs supports rerun consistency under fixed artifacts, not validity of the Full reference, robustness to resampling, or independent reproduction.

\begin{figure*}[t]\centering
\includegraphics[width=\textwidth]{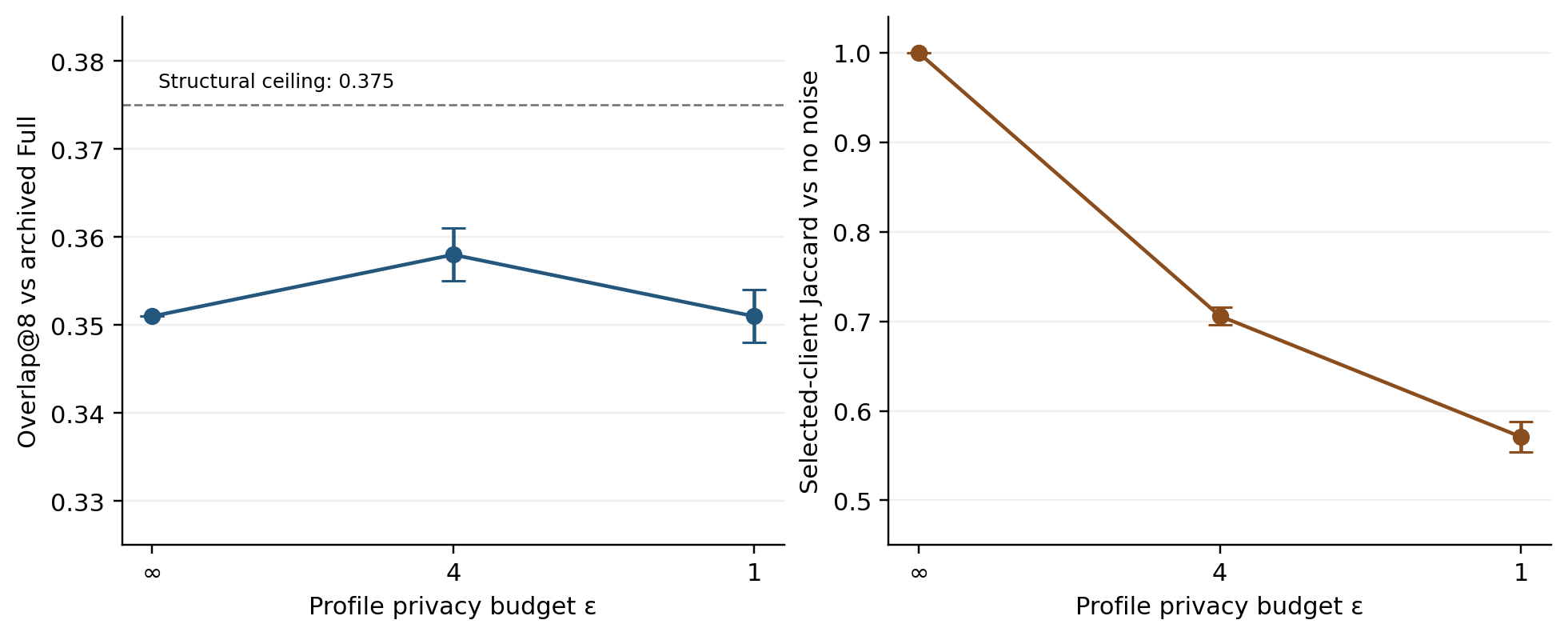}
\caption{Archived-reference evidence overlap and routing Jaccard across privacy budgets. Points and error bars use the published rounded seed-mean summaries and sample SDs in Table 4; they are not confidence intervals. The dashed line is the overlap ceiling in Eq. (13). The Full-reference mismatch remains unresolved.}\label{fig:2}
\end{figure*}

\subsection{Efficiency and Query-Sequence Behavior}

Figure 3 shows cumulative mean archived-reference overlap@8 for the four selective methods. Bands show the minimum and maximum across three seeds, not confidence intervals. The scorer is reinitialized for each dataset: the BioASQ segment starts a new online trajectory, although the plotted cumulative mean continues across the concatenated records. This figure does not demonstrate continuous adaptation across 200 questions, drift recovery, or recovery from client dropout. PACS-DP-3's lower archived end-to-end latency remains a descriptive observation. Without controlled stage-level and output-token comparisons, it cannot be attributed specifically to the routing score or latency penalty.

\begin{figure*}[t]\centering
\includegraphics[width=\textwidth]{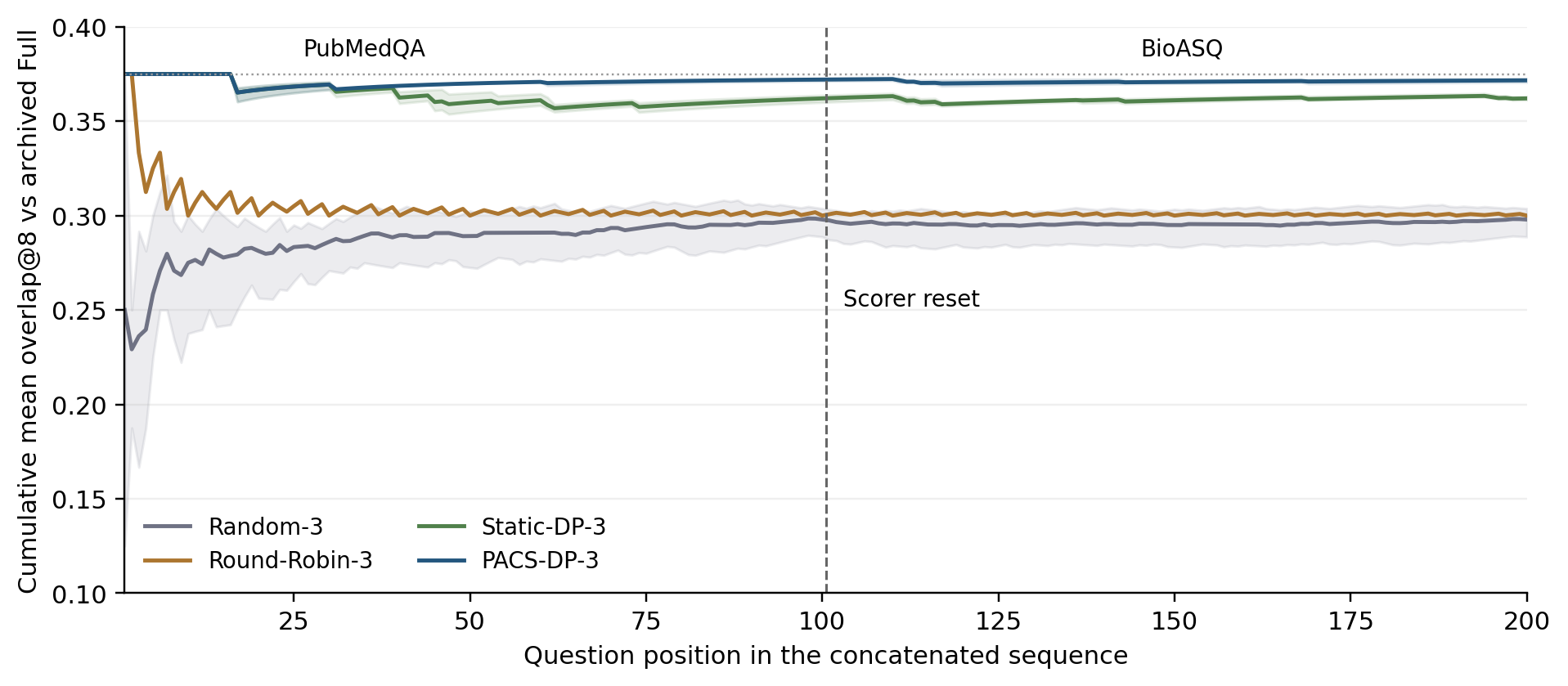}
\caption{Cumulative mean overlap@8 against archived Full, recomputed from the end-to-end per-query records. Shading is the min--max range across three seeds. The dashed vertical line marks the scorer reset between PubMedQA and BioASQ; the cumulative plotting statistic is not reset. The dotted horizontal line is 0.375.}\label{fig:3}
\end{figure*}

\subsection{Component Analysis}

\begin{table*}[t]\centering
\caption{Archived retrieval-only component analysis at \ensuremath{\varepsilon}=4: 200 questions \ensuremath{\times} three seeds; mean\ensuremath{\pm}sample SD (ddof=1). All variants contact three clients. The archived Full reference has incompatible tie-breaking. Generation is disabled.}\label{tab:5}
\small\setlength{\tabcolsep}{3pt}\renewcommand{\arraystretch}{1.12}
\newdimen\usablewidth\usablewidth=\dimexpr\textwidth-18pt\relax
\begin{tabular}{@{}>{\raggedright\arraybackslash}p{0.4364\usablewidth}>{\raggedright\arraybackslash}p{0.2182\usablewidth}>{\raggedright\arraybackslash}p{0.1909\usablewidth}>{\raggedright\arraybackslash}p{0.1545\usablewidth}@{}}\toprule

\textbf{Variant} & \textbf{Overlap@8 vs archived Full} & \textbf{Jain index} & \textbf{Total text bytes}\\\midrule

Full PACS-DP & 0.358\ensuremath{\pm}0.003 & 0.479\ensuremath{\pm}0.005 & 12,425\\

\ensuremath{-} historical utility (\ensuremath{\beta}=0) & 0.328\ensuremath{\pm}0.001 & 0.758\ensuremath{\pm}0.001 & 13,329\\

\ensuremath{-} explicit latency score (\ensuremath{\gamma}=0) & 0.367\ensuremath{\pm}0.006 & 0.420\ensuremath{\pm}0.001 & 10,538\\

\ensuremath{-} complete latency penalty (\ensuremath{\gamma}=0, \ensuremath{\tau}=0) & 0.375\ensuremath{\pm}0.000 & 0.360\ensuremath{\pm}0.000 & 9,913\\

\ensuremath{-} exploration (\ensuremath{\xi}=0) & 0.359\ensuremath{\pm}0.007 & 0.427\ensuremath{\pm}0.015 & 11,815\\

\bottomrule\end{tabular}
\end{table*}

Removing historical utility changes archived-reference overlap from 0.358 to 0.328 and increases participation uniformity from 0.479 to 0.758. Removing the complete latency penalty gives overlap 0.375, exactly the structural ceiling in Eq. (13), alongside lower Jain index and fewer text bytes. This does not demonstrate the best relevance or answer quality: the metric rewards avoiding two arbitrarily excluded reference clients. These ablations describe selection and payload behavior under one calibration and query order. They do not isolate a causal speed benefit of the latency penalty, and a common-protocol Full comparison is still required. Any change to fusion that changes the reward requires rerunning complete online trajectories rather than replacing final evidence in stored rows.

\subsection{Descriptive Analysis and Reproducibility}

No confirmatory significance or noninferiority claim is made. Repeated questions across seeds and online state dependence preclude treating all records as independent observations. Table 3 preserves its historical population SD across three seed means (ddof=0); Tables 4 and 5 preserve sample SD (ddof=1). These measures are descriptive and are not confidence intervals. Matching deterministic fields in two stored retrieval-only runs demonstrates consistency for fixed inputs; wall-clock fields remain observational. The analysis uses archived records; no new retrieval or generation experiment is reported. Independent reproduction requires the upstream data, indexes, models, and recorded environment.

\section{Discussion}

PACS-FedRAG implements a lightweight retrieval-budget decision using semantic profiles and four scalar statistics per client. The published records permit inspection of client selection, label predictions, calibration, and payload accounting. However, the principal evidence-overlap result is dominated by the interaction of disjoint partitions and deterministic RRF ties. The retrieval-only comparisons additionally reuse a Full reference from a different fusion protocol. Consequently, the archived results do not establish that adaptive feedback improves relevant evidence retrieval or preserves answer quality. Profile-noise and component changes demonstrably affect selected-client identities and participation concentration; the downstream value of those changes remains unresolved.

Profile protection, query-recipient reduction, and passage confidentiality are separate properties. The retrieval-only analytic Gaussian calibration concerns one bounded indexed record's contribution to a released centroid under fixed-size replace-one adjacency. Publicly seeded simulation does not supply a production DP guarantee. Fresh independently randomized releases need secure randomness and composition accounting, and the transcript containing passages and evidence feedback lies outside profile-only protection. Selecting three of ten clients reduces direct query recipients by 70\%, a property shared by all three-client selectors; it is not end-to-end confidential inference.

Feedback is sensitive to fusion semantics. With disjoint chunk identifiers, equal local ranks produce equal RRF scores. Insertion-order ties in the archived driver and document-ID ties in the retrieval-only driver select different subsets. Either deterministic convention can make a survival-based reward reflect an arbitrary ordering rather than relevance. Recomputing a same-protocol Full baseline repairs reference compatibility, but does not by itself repair the weak relevance interpretation of rank-only fusion in this partition. A relevance-sensitive fusion control and validated answer-level evaluation are needed. Malicious routing remains an untested integrity risk \cite{ref16}; exploration and EWMA are not demonstrated poisoning defenses.

The immediate next evaluation should use an identical, explicitly versioned fusion protocol for Full and selective methods, validate answer parsing with retained raw outputs, and include dataset-level and no-RAG controls. A changed fusion/reward policy requires new complete online sequences. Broader extensions include client-budget sweeps, topic-skewed partitions, query-order changes, availability changes, independently remeasured latency profiles, and distributed throughput and tail-latency measurements. Gold relevance judgments, dependence-aware uncertainty analysis, supervised-router comparisons, multi-centroid profiles, and malicious-client tests remain open work.

\section{Limitations}

The strongest limitation is measurement validity: archived-reference overlap is structurally capped and tied to arbitrarily favored clients, while the retrieval-only Full cache uses incompatible tie-breaking. Archived answer accuracy is below the fixed-label diagnostic, and absent raw responses prevent retrospective verification of parser behavior. Corrected same-protocol and answer-level results are not yet available. External validity is further limited by a sequential ten-client simulation, two restricted corpus subsets, the first 100 questions per dataset, one fixed budget, coarse hash partitions, fixed query order, and one latency calibration. The 600 question--seed records contain only 200 distinct questions. The scorer resets between datasets. Retrieval-only runs omit generation, and the earlier end-to-end run lacks equivalent analytic profile-calibration provenance. Neither three deterministic seeds nor a duplicate run demonstrates sensitivity to resampling or fresh production randomness. The study does not test malicious clients, profile poisoning, query privacy, passage confidentiality, dropouts, network contention, or composition across repeated releases.

\section{Conclusion}

PACS-FedRAG is an online fixed-budget selector that combines perturbed semantic profiles with evidence and systems feedback. Its implementation and archived results make the policy inspectable, and the three-client budget reduces direct query recipients by 70\% in the ten-client simulation. The archived end-to-end result is 0.372\ensuremath{\pm}0.001 overlap@8 against Full and 0.287\ensuremath{\pm}0.002 answer accuracy. Those values do not establish quality preservation: the overlap metric has a 0.375 structural ceiling, accuracy is below a fixed-label diagnostic, and the retrieval-only comparisons use a fusion-incompatible archived Full reference. Analytic Gaussian provenance supports the stated retrieval-only calibration, but not a production-private deployment or a retrospective guarantee for the earlier run. Corrected baselines, validated generation outputs, and complete reruns after reward-affecting changes are required before stronger performance conclusions can be drawn.

The implementation and versioned result records are available in the project repository. Reproducibility details are provided in Data and Code Availability.

\section*{Data and Code Availability}

\subsection*{Data availability}

The project repository at \url{https://github.com/alishams23/pacs-fedrag} contains the federation manifest, archived per-query records, summaries, analysis scripts, and retrieval-only artifacts under results/retrieval\_only\_v6. The Table 3 source is results/per\_query\_results.csv. The biomedical corpora and benchmark questions are obtained through the upstream Flower FedRAG/MIRAGE workflow and remain subject to their original licenses and access terms. The repository does not supply every corpus/index/model artifact required for an independent full rerun. Archived raw generator text is unavailable.

\subsection*{Code availability}

Code, experiment configurations, and archived result records are available at \url{https://github.com/alishams23/pacs-fedrag}. Model revisions, calibration settings, and artifact checksums are recorded in the accompanying run manifests. Reproducing the end-to-end benchmark additionally requires the documented Qwen2.5-7B-Instruct Q4\_K\_M generator environment.

\end{document}